\documentclass[twocolumn,showpacs,preprintnumbers,amsmath,amssymb,aps,prl]{revtex4}
\usepackage{graphicx}
\begin{document}

\title{Collective Sliding States for Colloidal Molecular Crystals}   
\author{C. Reichhardt and    
 C. J. Olson Reichhardt} 
\affiliation{ Theoretical Division, Los Alamos National Laboratory, 
Los Alamos, New Mexico 87545} 

\date{\today}
\begin{abstract}
We study the driving of
colloidal molecular crystals over
periodic substrates such as those created with optical traps. 
The $n$-merization that occurs in the colloidal molecular crystal states 
produces a remarkably rich variety of 
distinct dynamical behaviors,  
including polarization effects within the pinned phase 
and the formation of both ordered and disordered sliding phases.         
Using computer simulations, we map 
the dynamic phase diagrams as a function of substrate strength for 
dimers and trimers on a triangular substrate, and correlate features
on the phase diagram with transport signatures.  
\end{abstract}
\maketitle

\vskip2pc
A wide range of condensed matter systems can be modeled
as a collection of interacting particles on a periodic substrate
under equilibrium and nonequilibrium conditions, 
where the number of particles may be commensurate or incommensurate
with the substrate.
Examples of such 
systems include molecules and atoms on corrugated surfaces 
\cite{Coppersmith,Coppersmith2}, 
models of sliding interfaces \cite{Jensen}, 
vortices interacting with artificial pinning arrays in
 superconductors \cite{Baert,Harada,Reichhardt}, and  
vortices in Bose-Einstein condensates interacting with optical trap 
arrays \cite{Tung}. 
Recently, there has been growing interest in 
studying colloidal particles interacting 
with one-dimensional \cite{1D} or two-dimensional
periodic or quasiperiodic substrates such as those generated with optical traps.
\cite{Grier,Roth}.
Sufficiently strong traps can capture multiple colloids, 
and at commensurate fillings where the number of
colloids is equal to an integer multiple of the number of traps,
the colloids in each trap undergo an effective 
$n$-merization which causes them to
act like rigid objects such as dimers or trimers.
This produces an orientational degree of  
freedom and leads to the formation of ordered states which have 
been termed colloidal molecular crystals 
\cite{Olson,Brunner,Trizac,Frey}. 
Simulations and experiments have shown that these systems can 
undergo interesting ordered to orientationally ordered to disordered 
transitions 
as a function of substrate strength \cite{Olson,Brunner}. 
Additional theoretical and numerical studies demonstrated that 
colloidal molecular crystal 
systems can 
exhibit a rich variety of equilibrium states which 
have ferromagnetic, antiferromagnetic, 
and other types of spinlike symmetries \cite{Trizac,Frey}. 

Although individual colloids have been driven over periodic substrates
\cite{Korda,Spalding}, and colloidal assemblies have been depinned
from random substrates \cite{Ling},   
the nonequilibrium dynamics of colloidal
molecular crystals in the presence of an additional driving force 
has not been studied previously. The $n$-merization that occurs
in the colloidal molecular crystal system 
could be expected to produce dynamical sliding behaviors that are 
distinct from those of sliding point particles. 
In this work, we show that
the depinning force passes through a series of peaks as a function of
colloid density that are
associated with the formation of commensurate
colloidal molecular crystal 
states. 
At the commensurate densities, the scaling of the depinning threshold indicates
that the pinning is collective, while
pronounced changes occur in the depinning threshold as a function of substrate 
strength that correlate with changes in  the symmetry of the 
colloidal molecular crystal 
states. 
Changes in the 
structure 
of the colloidal molecular crystals 
also produce
features in the velocity-force curves.
We specifically map the dynamical phase diagram for 
dimer and trimer states at filling fractions of $2$ and $3$
colloids per trap. 
In addition to providing an understanding of the general 
phenomenon of sliding phases on periodic substrates, our results could
also be of importance for the development of externally driven dynamical
assembly techniques for colloids and other particulate matter systems.

We consider a two-dimensional 
system with periodic boundary conditions in the $x$ and $y$ directions
containing $N_{c}$ colloidal particles. The dynamics of a single colloid $i$
at position ${\bf R}_i$
is governed by the overdamped equation of motion \cite{Reichhardt}\
\begin{equation}
\eta \frac{d{\bf R}_i}{dt} = {\bf F}^{cc}_{i} + {\bf F}^{s}_{i} + {\bf F}_{d} ,
\end{equation}
where we set $\eta=1$.
The colloid-colloid interaction force is 
${\bf F}^{cc}_{i} = -\sum_{j\neq i}^{N_{c}} \nabla V(R_{ij})$ where
the potential has a Yukawa form, 
$V(R_{ij}) = (E_{0}/R_{ij})\exp(-\kappa R_{ij})$ and where
$R_{ij}=|{\bf R}_i-{\bf R}_j|$, $E_0=Z^{*2}/(4\pi\epsilon\epsilon_oa_o)$,
$\epsilon$ is the solvent dielectric constant, 
$Z^{*}$ is the effective charge, and $1/\kappa$ is the 
screening length. Lengths are measured in units of 
$a_{0}$, assumed to be on the order of a micron, 
forces are measured in units of $F_{0} = E_{0}/a_{0}$, 
and time is measured in units of $\tau = \eta/E_{0}$.   
The substrate force ${\bf F}_{s}$ 
arises from a triangular substrate with 
\begin{equation}
{\bf F}_{s} = \sum^{3}_{i=1}A\sin\left(\frac{2\pi b_{i}}{a_{0}}\right)
[\cos(\theta_{i}){\hat {\bf x}} - \sin(\theta_{i}){\hat {\bf y}}],
\end{equation}
where $b_{i} = x \cos(\theta_{i}) - y\sin(\theta_{i}) + a_{0}/2$, 
$\theta_{1} = \pi/6, \theta_{2} = \pi/2$,
and $\theta_{3} = 5\pi/6$. Here $A$ is the relative substrate strength and there are $N_{s}$ substrate minima.   
The initial colloidal positions are obtained through simulated annealing.
The applied driving force ${\bf F}_{d}=F_d{\bf \hat x}$ represents the force that
would be produced by an external electric field \cite{Ling}.    
For each drive, we measure the average colloid velocity 
$V=\langle N_c^{-1}\sum_i^{N_c}d{\bf R}_i/dt \cdot {\bf \hat x}\rangle$.
The depinning force $F_c$ is defined as the value of $F_d$ at
which $V=0.025$.

\begin{figure}
\includegraphics[width=3.5in]{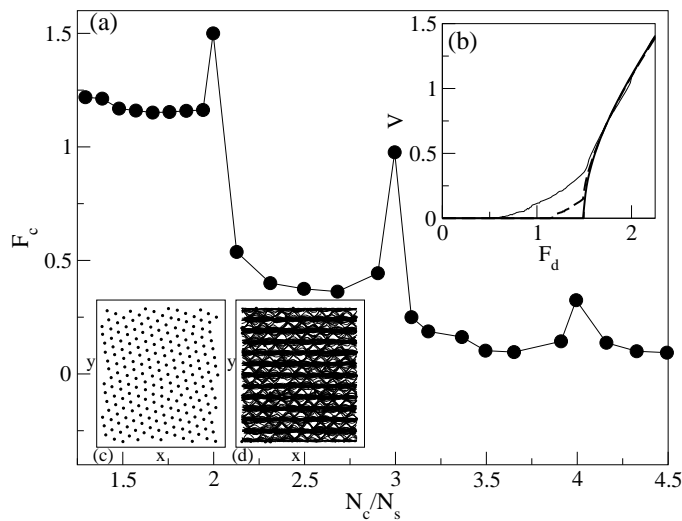}
\caption{
(a) The critical depinning force $F_{c}$ versus filling fraction
$N_{c}/N_{s}$ for a 
substrate with strength $A  = 2.5$. Peaks appear
at the commensurate fillings of $N_{c}/N_{s} = 2$, $3$ and $4$. 
(b) $V$, the average colloidal velocity per particle, 
vs the external drive $F_{d}$ for $N_{c}/N_{s} = 1.94$ (dotted line), 
$2$ (thick line), and $2.13$ (thin line). 
A clear depinning threshold exists which is largest
at $N_{c}/N_{s} = 2$.
(c) The colloidal positions (black dots) and trajectories (black lines) 
for $N_{c}/N_{s} = 2.0$ in
the pinned triangular crystal (PC) state at $F_{d} = 0.0$ and $A = 0.5$.
(d) Colloid positions and trajectories for $N_c/N_s=2.0$ in the 
moving random (MR) state at $A = 1.5$ and $F_{d}/F_{c} = 1.1$. 
}
\label{fig:depin}
\end{figure}

In Fig.~\ref{fig:depin}(a)
we plot the depinning threshold $F_{c}$ versus the filling factor $N_c/N_s$
over the range $1.5 < N_{c}/N_{s} < 4.5$
for a system
with a substrate of strength $A = 2.5$. 
At the integer matching fields $N_c/N_s=2$, 3, and 4,
there are clear peaks in the
depinning threshold.  This is similar to the behavior 
observed for vortices in superconductors
with periodic pinning, 
where peaks in the critical current 
(which is proportional to the depinning force)
appear when the vortex density is an integer multiple of the
pinning site density \cite{Baert,Reichhardt}.      
In Fig.~\ref{fig:depin}(b) we illustrate the velocity force curves 
for $N_{c}/N_{s} =  1.94$, $2.0$, and $2.13$. 
A clear single depinning threshold appears at $N_c/N_s=2$, 
while two-step depinning transitions occur at the 
noninteger fillings.
For $N_c/N_s=1.94$, the initial depinning occurs due to the 
motion of monomer defects in the dimer lattice followed by the depinning
of the remaining dimers, 
while for $N_{c}/N_{s} = 2.13$, the trimer defects in the dimer lattice
depin first and then the remaining dimers depin
at a higher drive. 

\begin{figure}
\includegraphics[width=3.5in]{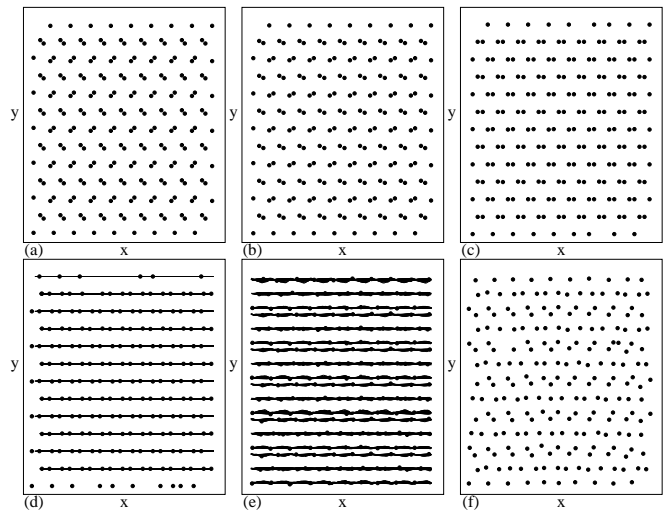}
\caption{
The colloid positions (black dots) and trajectories (black lines) 
for $N_{c}/N_{s} = 2.0$ and (a-d) $A = 3.25$; (e,f) $A=1.5$.
(a) The pinned herringbone (PHB) state at $F_{d} = 0.$
(b) The PHB at finite $F_{d}/F_{c} = 0.45$ 
showing the onset of dimer polarization in the direction of the drive.
(c) The fully polarized dimers at $F_d/F_c=0.9$ form the pinned 
ferromagnetic (PF) state. (d) The
moving ferromagnetic (MF) state at 
$F_{d}/F_{c} = 1.1$ where the motion
is strictly one-dimensional.   
(e) The PMF state for $A = 1.5$ at $F_{d}/F_{c} = 1.5$. 
(f) Vortex positions only in the PMF state from panel (e)
showing that every other row of dimers is aligned.
}
\label{fig:phb}
\end{figure}

\begin{figure} 
\includegraphics[width=3.5in]{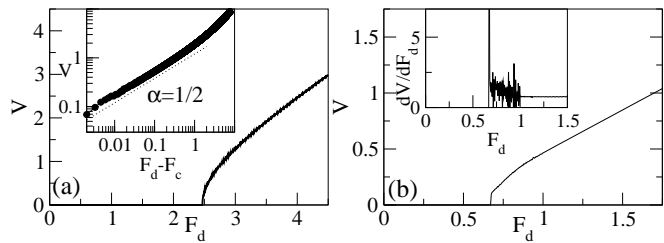} 
\caption{
(a) $V$ vs $F_{d}$ for $N_c/N_s=2.0$ and $A=3.25$
showing the 
continuous PF-MF depinning
transition. Inset: $V$ vs $F_{d} - F_{c}$ in the same
system indicating
a power-law scaling with $\alpha = 0.5$.
(b) $V$ vs $F_{d}$ for $N_{c}/N_{s} = 2.0$ and $A = 1.5$. 
The depinning occurs from a PHB state to a moving random (MR) state, 
and at higher drives the colloids organize into a partially-ordered 
moving ferromagnetic (PMF) state. 
Inset: $dV/dF_{d}$ vs $F_{d}$ for the same system.  
The sharp jump indicates that the
depinning transition is discontinuous. 
The fluctuating region corresponds to the MR state and
the onset of a regime with small fluctuations at $F_d>1$ corresponds to
the formation of the PMF state.
}
\label{fig:power}
\end{figure}

We measure $F_c$ versus substrate strength $A$ for 
the case $N_{c}/N_{s} = 2.0$ 
and find that there are three distinct pinned states and four moving states. 
In Fig.~\ref{fig:phb} we plot the colloidal configurations at different
values of $F_d$ for $A = 3.25$. 
At $F_{d} = 0$, the ground state is the
pinned herringbone (PHB) 
structure shown in Fig.~\ref{fig:phb}(a).
For increasing $F_d$, the dimers become increasingly {\it polarized} in the
direction of drive, as illustrated in Fig.~\ref{fig:phb}(b) for
$F_{d}/F_{c} = 0.45$. 
At sufficiently large $F_d$, 
the dimers are completely aligned into the
pinned ferromagnetic (PF) state shown in Fig.~\ref{fig:phb}(c) 
for $F_{d}/F_{c} = 0.9$. 
The system depins directly into the moving ferromagnetic (MF) state, where
the colloids move in one-dimensional channels as 
seen in Fig.~\ref{fig:phb}(d). 
This same sequence of states also appears for higher values of $A$. 
The PF-MF transition 
is continuous, as shown in Fig.~\ref{fig:power}(a) 
where we plot $V$ versus $F_{d}$ for $A = 3.25$.
The inset of Fig.~\ref{fig:power}(a) indicates that the velocity 
follows a power law scaling, $V = (F_{d} -F_{c})^{\alpha}$ with 
$\alpha = 1/2$, with a turnover at higher $F_{d}$ to a linear form. 
This exponent is consistent with elastic depinning, in which
the colloids keep the same neighbors while moving. 
We observe the same velocity scaling for $A \geq 2.5$.

For $0.7 < A < 2.5$, 
the PF state does not form. 
Instead, the PHB depins discontinuously 
into a {\it plastically} flowing state where the colloids
can exchange neighbors and undergo a transverse diffusive behavior
shown in Fig.~\ref{fig:depin}(d). 
We call this the moving random (MR) state.
In Fig.~\ref{fig:power}(b) we plot the velocity force curve
at $A = 1.5$ which has a discontinuous depinning transition, as indicated
in the inset of Fig.~\ref{fig:power}(b) where we show $dV/dF_d$ versus $F_d$.
There is a clear sharp jump in $V$ at the PHB-MR depinning transition,
followed by a regime of fluctuating $dV/dF_d$ which corresponds to the 
MR state. 
For  $F_{d} > 1.0$, the fluctuations are diminished when the system forms     
a partially-ordered moving ferromagnetic (PMF) state, illustrated in
Fig.~\ref{fig:phb}(e) and (f). 
Here the moving colloids form a stripelike structure 
in which every other row of dimers is aligned with the drive. 
Unlike the MR state shown in Fig.~\ref{fig:depin}(d), in the PMF state 
there is no transverse diffusion. 
For $A < 0.7$ and $F_{d} = 0$, the weak substrate causes
the elastic energy cost of the PHB state to be too high, and 
instead the colloids form a pinned triangular crystal (PC) state     
illustrated in Fig.~\ref{fig:depin}(c). 
The PC state is only weakly pinned 
and the depinning occurs as a continuous 
elastic transition to a moving triangular crystal (MC) state.

\begin{figure}
\includegraphics[width=3.5in]{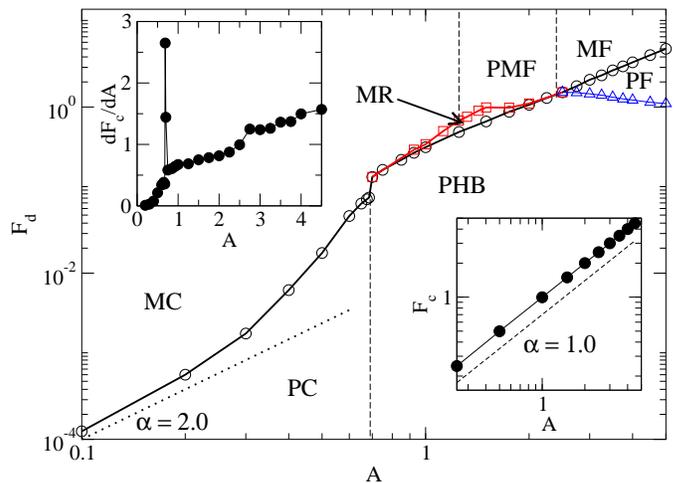}
\caption{
The dynamic phase diagram for $F_{d}$ vs $A$ with $N_{c}/N_{s} = 2.0$. 
Open circles: depinning threshold.
Dashed line: a fit to a power law with $\alpha = 2.0$. 
PC, pinned triangular crystal; PHB, pinned herringbone; 
PF, pinned ferromagnetic;
MC, moving triangular crystal; MR, moving random;
MF, moving ferromagnetic; and PMF, partially-ordered moving ferromagnetic.
Upper left inset: $dF_{c}/dA$ vs $A$ for the depinning curve
in the main panel.  The sharp peak separates the PC and PHB states. 
Lower right inset: $F_{c}$ vs $A$ for a single isolated particle, 
showing a linear scaling.
}
\label{fig:phase}
\end{figure}

In Fig.~\ref{fig:phase} we summarize the different states in a
dynamic phase diagram for $N_{c}/N_{s} = 2.0$.
We plot $dF_c/dA$ versus $A$ in the upper inset of Fig.~\ref{fig:phase} to show 
how changes in $F_{c}$ correlate with different phases of the system. 
The PHB-PC transition is marked by a discontinuous jump down in $F_c$ when
the PC phase forms, as indicated
by the sharp peak in $dF_{c}/dA$.
A smaller feature in 
$dF_{c}/dA$ occurs at $A=2.5$ when the
depinning changes from the discontinuous PHB-PMF transition to the
continuous PF-MF  transition.
The PF phase forms for $A > 2.5$ 
and the PHB-PF transition shifts to lower $F_d$ with increasing $A$. 
Since the PHB state forms due to the effective quadrupole interaction
between the dimers \cite{Trizac}, 
as $A$ increases, the colloids forming each dimer are pulled closer together,
reducing the quadrupole moment and facilitating the formation of the PF and
MF states.
The MR state appears
in a narrow window 
between the two elastic depinning transitions, PHB-MC and PHB-PMF.
In the MR regime, 
there is a competition between the quadrupole moment, 
which prevents the dimers from aligning into the
PF state, and the tendency of the external drive 
to align the dimers. 
For $ 0.7 < A \leq 1.25$, 
the MR state orders into a MC state for increasing $F_d$
rather than forming the PMF state.
For small $A$, $F_{c}$ exhibits a scaling $F_c \propto A^{2}$, 
consistent with collective
pinning in two dimensions \cite{Yu}. 
A similar scaling occurs for $A>0.7$ as well.  
In contrast, for a single colloid moving over the
periodic substrate, we find the linear behavior
$F_{c} \propto A$, as shown in the lower inset of Fig.~\ref{fig:phase}.
The collective pinning behavior that occurs in the PHB state for large $A$ 
likely arises because the objects that are 
depinning are dimers rather than single particles.     

\begin{figure}
\includegraphics[width=3.5in]{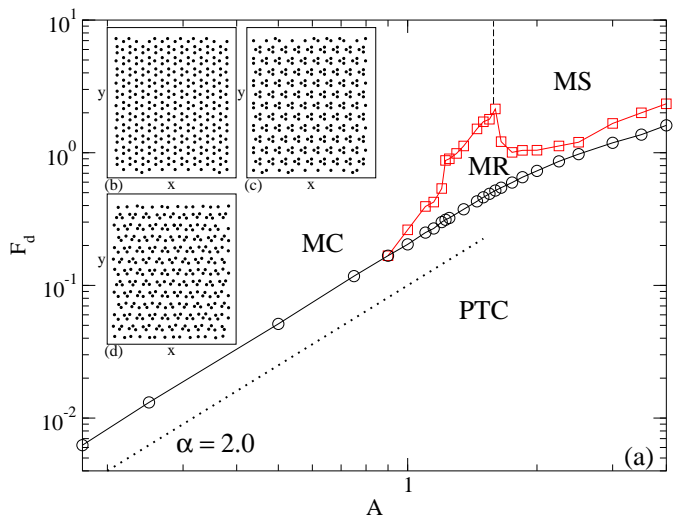}
\caption{
(a) The dynamic phase diagram for $F_{d}$ vs $A$ with $N_{c}/N_{s} = 3.0$. 
Open circles: depinning threshold.
PTC: pinned trimer crystal;
MC: moving triangular crystal;
MR: moving random;
MS: moving stripe.
(b) Colloid positions in PTC at $A=0.5$ and $F_d=0$.
(c) Colloid positions in PTC at $A=2.5$ and $F_d=0$. The size of
each trimer is reduced compared to panel (b).
(d) The MS state at $A=2.5$ and $F_d=1.5$.
}
\label{fig:phasetri}
\end{figure}
         
For $N_{c}/N_{s} = 3.0$, we find a similar set of pinned and moving phases 
as shown in the dynamic phase diagram for $F_{d}$ vs $A$ in 
Fig.~\ref{fig:phasetri}(a). 
At this filling, there is only a single pinned phase, the pinned trimer
crystal (PTC) illustrated in Figs.~\ref{fig:phasetri}(b) and (c) for
$A = 0.5$ and $A = 2.5$,
and so there is no discontinuous transition
in $F_{c}$ vs $A$ such as that seen at the PHB-PC transition
for  $N_{c}/N_{s} = 2.0$.
In Fig.~\ref{fig:phasetri}(a), the PTC-MC depinning transition for $A<1.0$ is
elastic, while
for $ A > 1.0$ the PTC-MR depinning transition is followed at higher
drives by the organization of the colloids into either the moving crystal (MC)
state or into the moving stripe (MS) state illustrated in 
Fig.~\ref{fig:phasetri}(d).
In the MS state, the colloids are strictly confined to move along 
one-dimensional rows with no transverse diffusion. 
The oriented trimer structure found in the PC is lost 
in the MS state; however, some orientational ordering of the trimers
persists in the MS state, producing a zig-zag structure.
The MR state reaches its maximum extent at the transition
between the MC and MS phases at $A=1.6$.   
The scaling of 
$F_c$ with $A$ is consistent
with collective depinning for low $A$, while for $A > 3.0$, 
there is a rollover to a more linear regime, 
indicative of single particle depinning. 
In the high-$A$ regime, the depinning occurs via the hopping of
individual colloids from well to well, followed 
at higher drives by more general motion of
all the colloids. 

We expect that a similar type of phase diagram will 
occur for the higher order molecular crystals.
Additional phases are likely to occur 
at incommensurate fillings.  Thermal 
fluctuations could produce interesting results and new effects
since it has been shown that the
disordering transition depends on the substrate strength 
\cite{Olson,Brunner,Trizac}.   

In summary, we find that a remarkably rich variety of dynamical 
sliding states can be 
realized for colloidal molecular crystals  
under an external drive
with varied substrate strength. 
For dimer colloidal molecular crystals,
these states include moving ordered, partially ordered, and moving random
phases. The external drive can induce transitions 
within the pinned phase itself such
as a polarization from a pinned
herringbone to a pinned ferromagnetic state. 
The onset of the different phases can be identified through features 
in the transport response and depinning threshold.    
We map the dynamical phase diagrams for 
dimer and trimer states and find that similar features exist for 
both fillings.   
Our results should  be useful for the general understanding 
of sliding states for complex particles on periodic substrates 
and may also have potential applications for creating
dynamically induced self-assembled structures.

This work was carried out under the auspices of the 
NNSA of the 
U.S. DoE
at 
LANL
under Contract No.
DE-AC52-06NA25396.

\end{document}